\documentclass[12pt]{arXiv_class}
\usepackage{mathrsfs}

\title{\textbf{Few-Photon Heterodyne Spectroscopy}} 
\author{Gustavo C. Amaral, T. Ferreira da Silva, G. P. Tempor\~{a}o and J. P. von der Weid}

\begin{document}
\maketitle

\begin{abstract}
We perform a high resolution Fourier Transform Spectroscopy of optical sources in the few-photon regime based on the phenomenon of two-photon interference in a beam splitter. From the heterodyne interferogram between test and reference sources it is possible to obtain the spectrum of the test source relative to that of the reference. The method proves to be a useful asset for spectral characterization of faint optical sources below the range covered by classical heterodyne beating techniques.
\end{abstract}

First observed by Hanbury-Brown and Twiss in 1956 during stellar intensity interference measurements \cite{BrownNature1956}, two-Photon interference has been a much more familiar concept since the late eighties, when Hong, Ou and Mandel demonstrated that photons generated from Spontaneous Parametric Down Conversion (SPDC) tend to bunch together when directed into a symmetric optical beam splitter \cite{HOMPRL1987}. This experimental breakthrough deconstructed a strong view, advocated by Dirac, by which photons would only interfere with themselves \cite{DiracPQM4Ed1958}. In this context, the temporal correlation among photons is a major concern for multi-photon interference since, in fact, the photons do not interfere with each other but, rather, the wave packet that describes both photons interferes with itself \cite{OuBOOK}. When the relative phase delay between the two wave packets is matched within the photon's mutual coherence in the HOM Interferometer \cite{HOMPRL1987}, a destructive quantum interference effect is seen in the form of a ``dip" in the coincidence counting rate, the so-called \textit{HOM dip}. Even though it may seem, due to the nature of the HOM experiment, that the interference is generated owing to a somewhat local interaction between two single photons, this is not exactly the case. Two-photon interference between photons arriving at the beam splitter at much different times has been experimentally demonstrated, for which the picture of interference between two individual photons is not applicable \cite{PittmanPRL1996}.

After the first demonstration of the \textit{photon bunching} effect \cite{HOMPRL1987}, the phenomenon was also observed for weak coherent states, the highly attenuated light emerging from a conventional laser source \cite{RarityJOB2005,ThiagoJOSAB2015}. The higher probability of multi-photon emission in this case, however, limits the visibility to $50\%$ while, for SPDC states, the visibility may reach almost $100\%$ \cite{OuBOOK, GerritsPRA2015}. In more recent developments, it has been shown that frequency-displaced indistinguishable photons generate a beat pattern that modulates the interference curve \cite{LegeroAPB2003}. The resulting interferogram translates the spectral characteristics of the beating photons and, as it was shown in \cite{ThiagoJOSAB2015}, spectroscopy on the few-photon regime can be achieved. Even though mention is made to the possibility of arbitrary laser characterization, it has not been experimentally demonstrated. The reported proof-of-principle is constructed based on a self-heterodyne setup in which two weak coherent states are mutually characterized.

In this Letter, we develop the mathematical model which permits one to associate the Fourier Transform of the interferogram of the two-photon interference in a Hong-Ou-Mandel interferometer to the spectral shape of the interfering wave-packets and experimentally demonstrate the feasibility of high resolution spectral characterization of an arbitrary single-mode laser source in the few-photon regime. Furthermore, the limitations of the the few-photon spectral characterization and the region of operation in terms of the average number of photons of the sources being characterized are discussed.

\begin{figure}[H]
\centering
\fbox{\includegraphics[width=\linewidth]{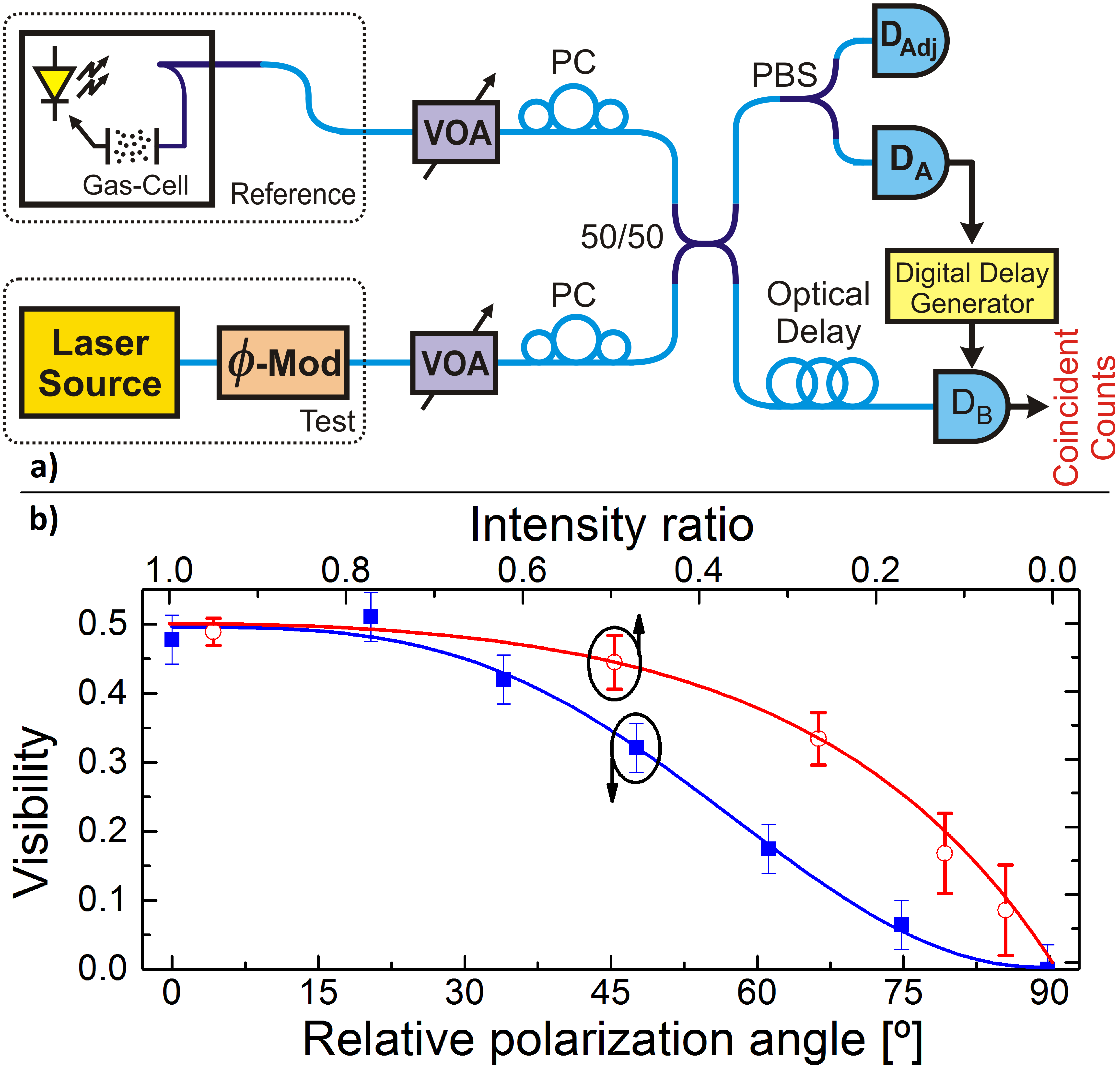}}
\caption{a) Experimental setup for the spectral characterization of optical sources in the few-photon regime. VOA: variable optical attenuator; PC: polarization controller; $\varphi$-Mod: phase modulator; PBS: polarization beam splitter; D$_n$: single-photon detector. b) Dependence of the interference visibility on the relative polarization and intensity ratio of the states: \textit{red line} visibility versus relative intensity ratio; \textit{blue line} visibility versus relative polarization angle.}
\label{fig:LaserCharacterizationSetup}
\end{figure}

The basic experimental apparatus for the spectral characterization in the few-photon -- or weak -- regime is shown in Fig. \ref{fig:LaserCharacterizationSetup}-a, where two independent lasers, \textit{reference} and \textit{test}, and single-mode fibers are employed. The \textit{reference} laser is an external cavity laser diode frequency-stabilized at 1547.32 nm by a high Q-factor gas cell. The attenuated signal from the \textit{reference} laser is directed to a HOM Interferometer along with the optical signal of the \textit{test} laser. The \textit{test} source is composed of a tunable wavelength laser, a 90/10 splitter, a second gas-cell, and a PID system connected to the wavelength tuning input of the laser. By controlling the region of the gas-cell's absorption line to which the PID system is locked, one can enforce the desired frequency-displacement between \textit{reference} and \textit{test} sources. This way, we could assess the validity of the Few-Photon FTS with a wide range of values for $f_{beat}$, the relative frequency displacement between coherent states. The input coherent states are power-balanced by variable optical attenuators as to reach the few-photon regime with the same average number of photons per pulse. Also, the counts in InGaAs-based Single-Photon Avalanche Photodiodes (SPADs) connected to a polarizing beam splitter are minimized so that the photons reaching the interferometer are polarization-aligned. Fig. \ref{fig:LaserCharacterizationSetup}-b shows the drop in the interferogram visibility as a function of the distinguishability of the states, which is governed by their relative states of polarization and relative intensities inside the time delay for which the temporal modes are overlapped in the interferometer since we consider the single-mode case. The bunching effect depends on the indistinguishability between wave-packets that describe the interaction between two photons and the beam splitter so, if the photons can be distinguished, the interference pattern is lost \cite{HOMPRL1987,ThiagoJOSAB2015}. The HOM Interferometer is completed by connecting the output spatial modes of the 50/50 beam splitter in two SPADs -- $D_A$ and $D_B$ -- operating in gated Geiger mode. $D_A$ triggers $D_B$ whenever a photon-counting event occurs therefore each count at $D_B$ corresponds to a coincidence event which, as a function of the relative delay between detectors, compose the HOM interferometer's visibility curve. This process of postselection guarantees that only the mode-matched photons contribute to the interferogram. An optical fiber loop is connected before $D_B$ to optically compensate the electrical delay between the detectors. Both detectors are set to $15\%$ efficiency and 4 ns gate width which corresponds to the best compromise between dark counts (on the order of $10^{-5}$ per detection gate) and efficiency.

Comparative classical heterodyne spectroscopic measurement can be performed by slightly modifying the setup of Fig. \ref{fig:LaserCharacterizationSetup}-a to overlap bright (non-attenuated) versions of \textit{test} and \textit{reference} lasers in a \textit{pin} photodiode. The frequency beat is visualized in an Electrical Spectrum Analyzer (ESA). The results of Fig. \ref{fig:LaserCharacterizationSetup}-b show that the visibility drop due to misalignment in the interferometer, even though a concern, is not extremely severe. Deviations of $20\%$ in both polarization alignment and intensity ratio can still be tolerated and fall well inside the SPAD's sensitivity. The fitting model for the relative intensity ratio is the visibility equation: $V\!=\!2R\!/\!\left(\!1\!+\!R\!\right)^2$ \cite{RarityJOB2005}, where $R$ is the average number of photons per gate ratio between the two sources and $V$ is the visibility. In the case of the relative state of polarization (SOP) alignment, the fitting model is obtained using Malus's Law \cite{SalehBOOK} applied to the the visibility equation. We consider the projection on the measurement basis to write $R\!=\!cos^2\!\left(\!\theta_r\!\right)\!/\!\left[\!1\!+\!\sin^2\!\left(\!\theta_r\!\right)\!\right]$, where $\theta_r$ is the relative polarization angle between the interfering beams. The maximum value of the visibility in the case presented in Fig. \ref{fig:LaserCharacterizationSetup}-b is $V_{max}\!=\!0.5$ due to the limitation associated to coherent states. The method is therefore robust regarding polarization, intensity and spatial-temporal mode-matching conditions.

Classically, the Interference Equation of two partially coherent waves has the form $I\!=\!I_1\!+\!I_2\!+\!2\!\sqrt{I_1 I_2}|g_{12}|\!\cos \varphi$ \cite{SalehBOOK}, where $I_1$ and $I_2$ are the respective intensities of each source, $g_{12}$ is their mutual coherence, and $\varphi$ is their relative phase difference. Writing the Interference Equation in terms of the wave's power spectral density, $S\left(\nu\right)$, yields $I\!\left(\!\tau\!\right)\!=\!2\!\int^{\infty}_0\!S\!\left(\!\nu\!\right)\!\left[\!1\!+\!\cos\!\left(\!2\!\pi\!\nu\!\tau\!\right)\!\right]\!d\!\nu$ if one assumes $I_1\!=\!I_2$ and where $\tau$ is the relative temporal delay between the two optical wave-packets. From this result, it is possible to determine the spectrogram of the beat of two light sources by taking the inverse-Fourier transform of the respective interferogram in a process known as \textit{Fourier-transform Spectroscopy} \cite{SalehBOOK}. The proposed Few-Photon Spectroscopy is an application of the Fourier-transform spectroscopy to light with low average photon flux taking advantage of two-photon interference in a Hong-Ou-Mandel Interferometer.

In the quantum mechanical context, the mutual coherence of the wave-packets ($g_{A,B}$) defines the joint detection probability between detectors $D_A$ and $D_B$ and has the following mathematical expression dependent on the temporal wave-packets describing \textit{reference} and \textit{test} states, $f_1\!\left(\!t\!\right)$ and $f_2\!\left(\!t\!\right)$ respectively \cite{LegeroAPB2003, WangJPB2006}:
\begin{equation}
g_{A,B}\!\left(\!t_0,\tau\!\right)\!=\!\tfrac{1}{4}\!\big|\!f_1\!\left(\!t_0\!+\!\tau\!\right)\!f_2\!\left(\!t_0\!\right)\!-\!f_1\!\left(\!t_0\!\right)\!f_2\!\left(\!t_0\!+\!\tau\!\right)\!\big|^2
\label{eq:gAB}
\end{equation}
One can manipulate Eq. \ref{eq:gAB} to find that
\begin{align}
\begin{aligned}
g_{A,B}\!\left(\!t_0,\tau\!\right)\!&=\!\tfrac{1}{4}\!\big|\!f_1\!\left(\!t_0\!\right)\!\big|^2\! \big|\!f_2\!\left(\!t_0\!+\!\tau\!\right)\!\big|^2\!+\!\big|\!f_1\!\left(\!t_0\!+\!\tau\!\right)\!\big|^2\!\big|\!f_2\!\left(\!t_0\!\right)\!\big|^2 \\
&\hspace{0.6cm}-\!\left[\!f_1\!\left(\!t_0\!\right)\!f_2\!\left(\!t_0\!+\!\tau\!\right)\!f_1^{\ast}\!\left(\!t_0\!+\!\tau\!\right)\!f_2^{\ast}\!\left(\!t_0\!\right)\!\right.\\
&\hspace{1.0cm}+\!\left.\!f_1\!\left(\!t_0\!+\!\tau\!\right)\!f_2\!\left(\!t_0\!\right)\!f_1^{\ast}\!\left(\!t_0\!\right)\!f_2^{\ast}\!\left(\!t_0\!+\!\tau\!\right)\!\right]
\end{aligned}
\label{eq:gAB_2}
\end{align}
where $^{\ast}$ denotes the complex conjugate. Now we define $r\!\left(\!t\!\right)\!=\!\big|\!f_1\!\left(\!t\!\right)\!\big|^2$, $s\!\left(\!t\!\right)\!=\!\big|\!f_2\!\left(\!t\!\right)\!\big|^2$, and $f_1\!\left(\!t\!\right)\!f_2^{\ast}\!\left(\!t\!\right)\!=\!z\!\left(\!t\!\right)$. To obtain the probability of detecting two photons in detectors $D_A$ and $D_B$ with a time difference of $\tau$ ($P_{c}\!\left(\!\tau\!\right)$), one integrates over all possible values of $t_0$, the arrival time of the photons, obtaining the curve that corresponds to the interferogram as a function of $\tau$. Upon integration over $t_0$, we find that $P_{c}$ has the following expression
\begin{align}
\begin{aligned}
P_{c}\!\left(\!\tau\!\right)\!&=\!\frac{1}{4}\!\left[\!\int_{-\infty}^{\infty}\!r\!\left(\!t_0\!\right)\!s\!\left(\!t_0\!+\!\tau\!\right)\!dt_0\!+\!\int_{-\infty}^{\infty}\!r\!\left(\!t_0\!+\!\tau\!\right)\!s\!\left(\!t_0\!\right)\!dt_0\right.\\
&\hspace{0.4cm}-\!\left(\!\int_{-\infty}^{\infty}\!z\!\left(\!t_0\!\right)\!z^{\ast}\!\left(\!t_0\!+\!\tau\!\right)\!dt_0\!+\!\left.\!\int_{-\infty}^{\infty}\!z^{\ast}\!\left(\!t_0\!\right)\!z\!\left(\!t_0\!+\!\tau\!\right)\!dt_0\right)\!\right]
\end{aligned}
\label{eq:P_Coinc}
\end{align}

Taking the Fourier Transform of Eq. \ref{eq:P_Coinc} yields
\begin{align}
\begin{aligned}
\mathscr{F}\!\left\{\!P_{c}\!\left(\!\tau\!\right)\!\right\}\!&=\!\tfrac{1}{4}\!\left(\!R\!\left(\!\omega\!\right)\!S\!\left(\!\omega\!\right)\!+\!R\!\left(\!\omega\!\right)\!S\!\left(\!\omega\!\right)\!\right.\\
&\hspace{0.6cm}-\!\left.\!\left[\!Z\!\left(\!\omega\!\right)\!Z^{\ast}\!\left(\!\omega\!\right)\!+\!Z^{\ast}\!\left(\!\omega\!\right)\!Z\!\left(\!\omega\!\right)\!\right]\!\right)
\end{aligned}
\label{eq:FourierP_Coinc}
\end{align}
Using properties of the Fourier Transform, and setting $\mathscr{F}\left\{f_i\left(t\right)\right\}=\phi_i\left(\omega\right)$, a somewhat extensive but straightforward calculation allows one to rewrite Eq. \ref{eq:FourierP_Coinc} as:
\begin{align}
\begin{aligned}
\mathscr{F}\!\left\{\!P_{c}\!\left(\!\tau\!\right)\!\right\}\!&=\!\tfrac{1}{2}\!\left[\!\phi_1\!\left(\!\omega\!\right)\!\ast\!\phi_1\!\left(\!\omega\!\right)\!\times\!\phi_2\!\left(\!\omega\!\right)\!\ast\!\phi_2\!\left(\!\omega\!\right)\!\right.\\
&\hspace{0.6cm}-\!\left.\!\phi_1\!\left(\!\omega\!\right)\!\ast\!\phi_2^{\ast}\!\left(\!\omega\!\right)\!\times\!\phi_1^{\ast}\!\left(\!\omega\!\right)\!\ast\!\phi_2\!\left(\!\omega\!\right)\!\right]
\end{aligned}
\label{eq:FinalP_Coinc}
\end{align}
Upon close examination of Eq. \ref{eq:FinalP_Coinc}, one distinguishes that the first term corresponds to the DC component of both $f_1\left(t\right)$ and $f_2\left(t\right)$, and also to a component centered at twice the central frequency of the optical carriers. This high-frequency component is filtered out since the detectors are not sufficiently broad-band to account for its detection whereas the DC component can be neglected since it doesn't contain any spectral information -- it is associated to the average number of coincidence events. The second term, however, corresponds to the spectral component centered at the beat frequency between the optical carrier frequencies shaped as the modulus squared of the convolution of the spectral shapes of the temporal wave-packets. This component is the frequency beat that can be visualized in the Electrical Spectrum Analyzer from which classical spectral characterization can be performed.

We find, therefore, that the spectral characterization of optical sources in the few-photon regime is performed by taking the Fourier Transform of the two-photon interferogram which yields the convoluted spectral shapes of the individual states centered at the beat frequency between the optical carriers, a result condensed by Eq. \ref{eq:Few-PhotFourierSpec_ModelEq} which defines the \textit{Few-Photon Fourier Transform Spectroscopy} (Few-Photon FTS).
\begin{equation}
\big|\phi_1\left(\omega\right) \ast \phi_2\left(\omega\right)\big|^2 \simeq \mathscr{F}\left\{\int_{-\infty}^{\infty} g_{A,B}\left(t_0,\tau\right) dt_0\right\}
\label{eq:Few-PhotFourierSpec_ModelEq}
\end{equation}
Even though the final result is in a convolution form, the process of deconvolution in order to determine the spectrum of \textit{test} source is straightforward since the spectral characteristics of the \textit{reference} source are known \textit{a priori} \cite{WienerBOOK1949}. It is worth noting that the \textit{test} source's intensity is related to the average number of photons impinging on the detector during a temporal gate window and, therefore, can be easily determined.

\begin{figure}[H]
\centering
\fbox{\includegraphics[width=\linewidth]{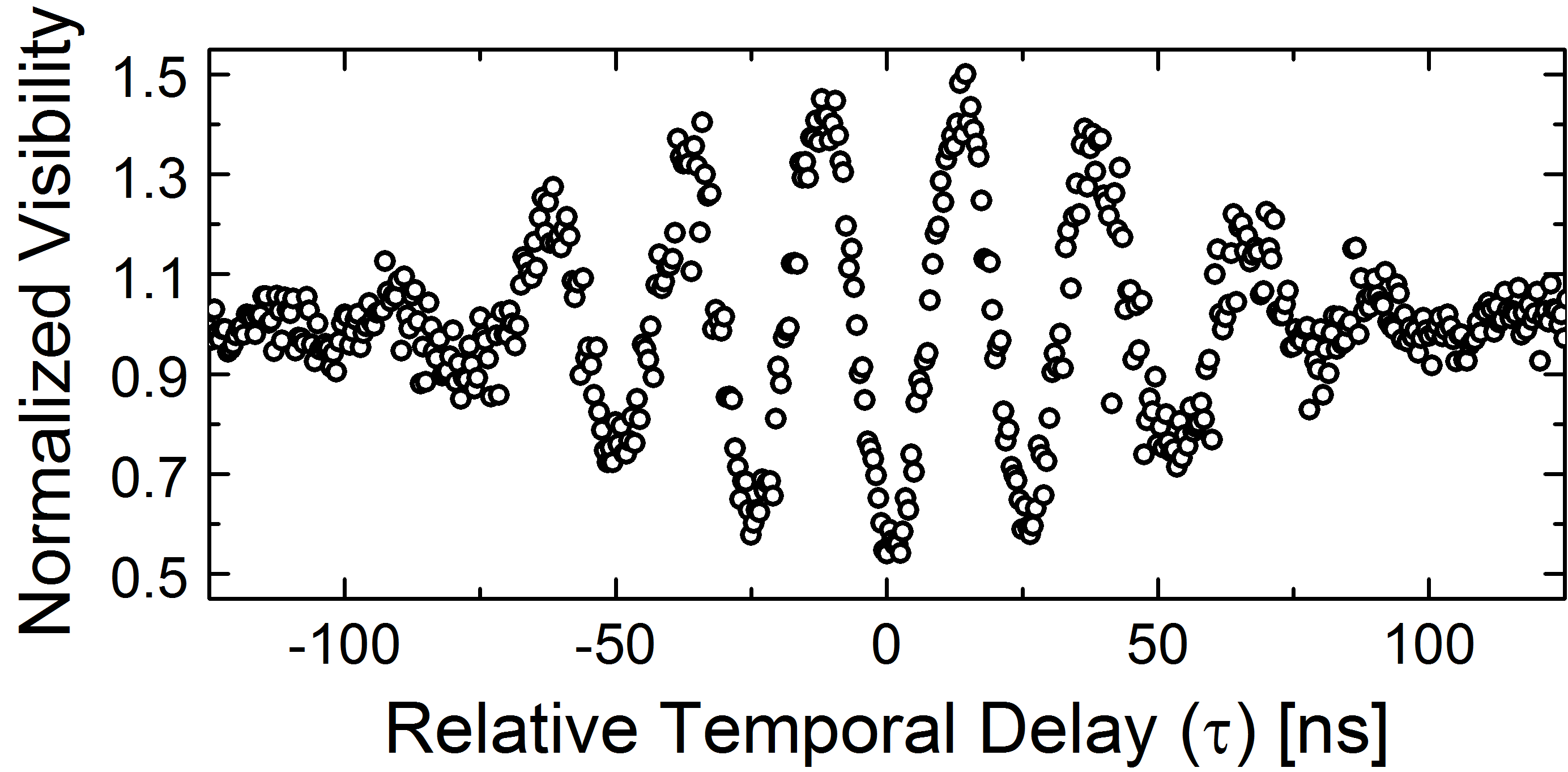}}
\caption{Interferogram of the two-photon HOM interference between \textit{test} and \textit{reference} weak-coherent states frequency-displaced by $40$ MHz with $\mu = 0.2$.}
\label{fig:TypicalFewPhotInterferogram}
\end{figure}

A typical interferogram curve is presented in Fig. \ref{fig:TypicalFewPhotInterferogram} for a frequency of $40$ MHz. The time delay span was chosen taking into account mutual temporal coherence of the wave-packets as to cover the whole region of overlapping temporal modes. The polarization states are aligned up to the detector's sensitivity level and the average number of photons per detection gate is set to $\mu = 0.2$. The interferogram presented in Fig. \ref{fig:TypicalFewPhotInterferogram} was acquired by sweeping the relative delay between the detectors in the HOM interferometer in steps of $500$ ps respecting the two-photon interference conditions described in Fig. \ref{fig:LaserCharacterizationSetup}-b and the discussion thereupon.

\begin{figure}[H]
\centering
\fbox{\includegraphics[width=\linewidth]{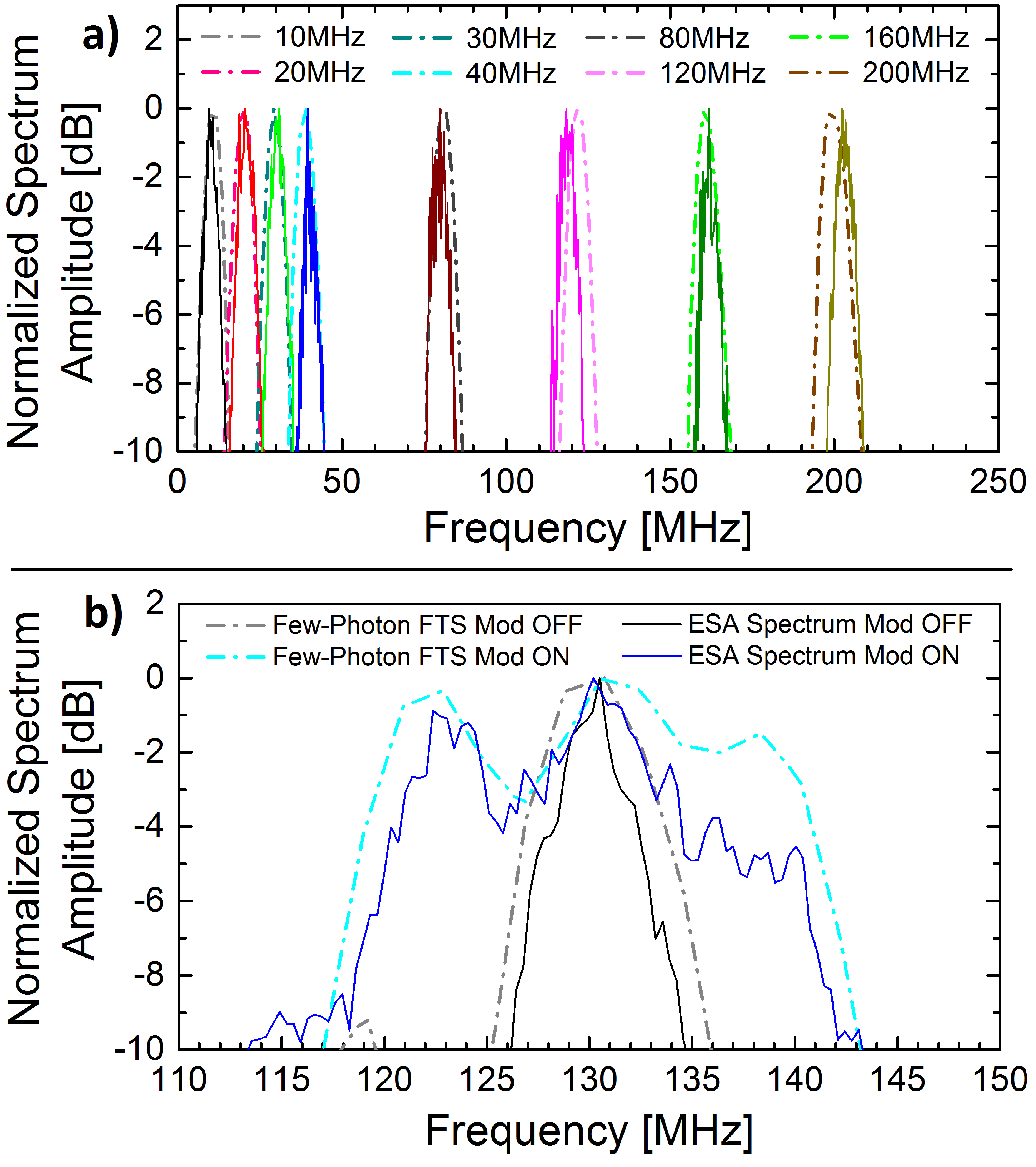}}
\caption{a) Comparison between the two spectral characterization techniques: Few-Photon FTS (dashed lines) and classical heterodyne beat (solid lines). The results show good agreement for a wide range of frequencies. b) The electro-optic phase modulator is switched on and off for a constant beat frequency of the lasers. The enlargement of the spectrum with respect to $f_{beat}$ is accurately described by the Few-Photon FTS.}
\label{fig:SpectrumComparison}
\end{figure}

Figure \ref{fig:SpectrumComparison}-a superposes the resulting spectral shapes of the Few-Photon FTS and those measured in the ESA for different values of $f_{beat}$, ranging from $10$ MHz to $200$ MHz, with good agreement. Slight relative shifts are due to the instability of the gas-cell that locks the frequency of the \textit{test} source. We also enforced non-gaussian spectra on the \textit{test} laser by means of a phase modulator at its output driven by an Arbitrary Waveform Generator (AWG) (refer to Fig. \ref{fig:LaserCharacterizationSetup}). By selecting the modulation parameters, the spectrum can be widened, which reflects on the temporal narrowing of the HOM interferogram due to the inverse relation between the mutual temporal coherence ($\tau_c$) and the spectral linewidth ($\Delta \nu$) \cite{SalehBOOK}. Such spectral widening is restored from the interferogram when we employ the Fourier Transform and is presented in Fig. \ref{fig:SpectrumComparison}-b where, once again, we compare the results of the Few-Photon FTS and of the classical heterodyne beat.

Note that, due to the relative nature of the method and the folding nature of the FFT spectrum, there is an ambiguity in determining the \textit{test} source's center frequency which can lie on $f_{ref}\!\pm\!f_{beat}$. Running the method twice while displacing the reference frequency by an amount smaller than $f_{beat}$ eliminates this ambiguity. Considering the deconvolution process, it is also worth noting that a proper choice of the reference frequency must result in an unfolded frequency spectrum, so that $f_{beat}$ should be larger than the width of the \textit{test} source spectrum \cite{WienerBOOK1949}.

\begin{figure}[H]
\centering
\fbox{\includegraphics[width=\linewidth]{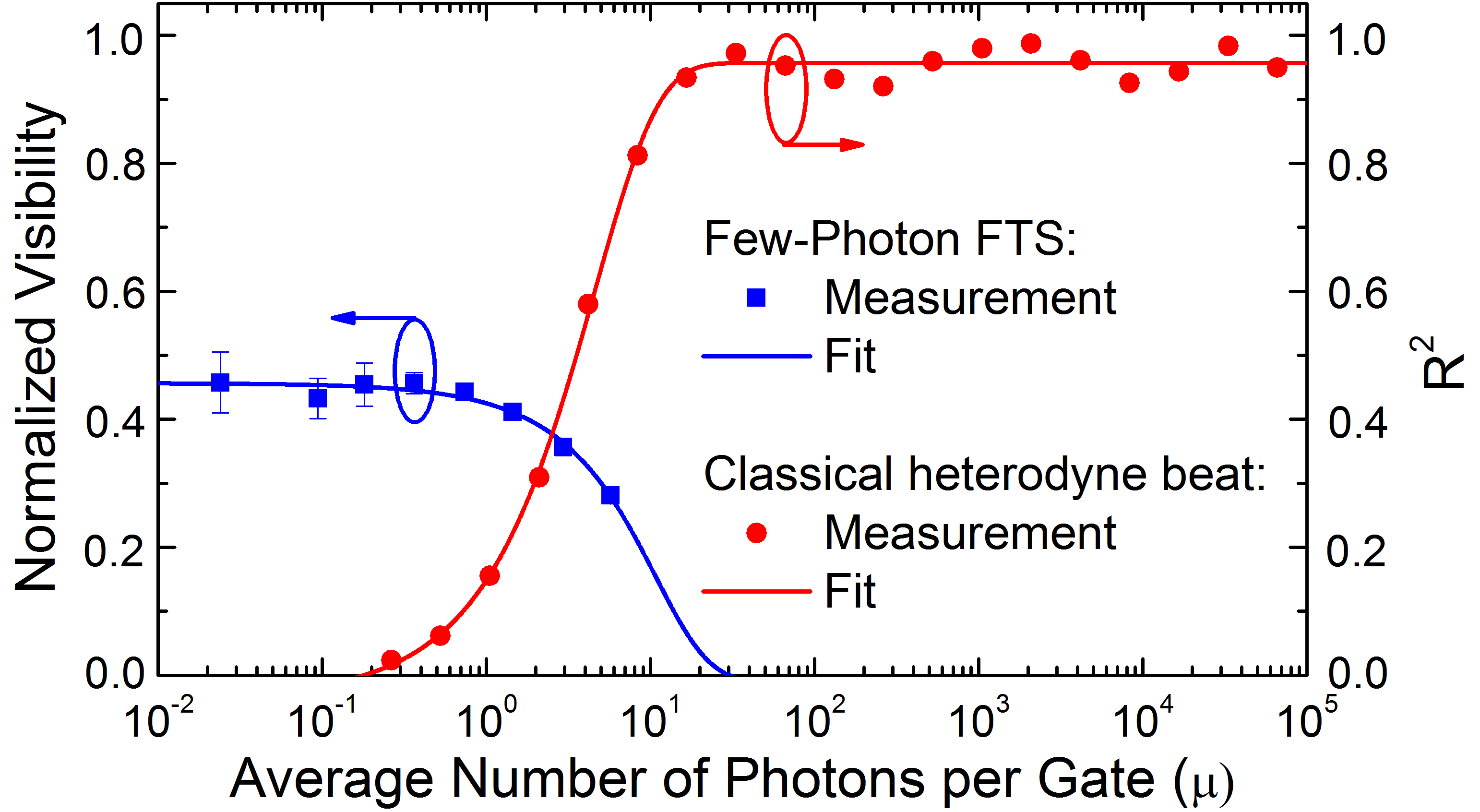}}
\caption{Range of application of the Few-Photon FTS and classical spectroscopy methods in terms of the average number of photons per pulse, $\mu$, of the \textit{test} source. The $R^2$ parameter of the approximated model and the visibility of the HOM interferometer are set as parameters of the accuracy and, therefore, the validity of each technique. The data sets were fit to evidence the dependency of the effectiveness on $\mu$. Maximum $P_{ref}$ is $+13$dBm.}
\label{fig:EffectivenessRegion}
\end{figure}

Since both techniques are shown to be equivalent both by the mathematical model and by the experimental results, a distinction has to be made between them. This is translated in the region of effectiveness of each method, the Few-Photon FTS figuring as a better candidate as soon as the average number of photons per pulse falls below a certain value. Fig. \ref{fig:EffectivenessRegion} displays the region of effectiveness of the classical heterodyne beat and the Few-Photon FTS as a function of the average number of photons per gate of the \textit{test} source. The employed criterion of effectiveness for the Few-Photon FTS is the Visibility of the interferogram curve since no spectral analysis can be performed once this value goes to zero. In the case of the classical heterodyne beat, the data acquired in the ESA was approximated to a simple model when one assumes gaussian wave-packets for the interfering beams, i.e., a gaussian-shaped spectral curve centered at the beat frequency. The $R^2$ (R-squared) parameter of the approximation, which translates the likelihood between the model and the acquired data, was used as a criterion for the technique's effectiveness.

Throughout all measurements, the relative intensity level and polarization angle was kept at the optimal value to guarantee maximum visibility in the interference pattern, i.e., $\mu_{ref} = \mu_{test}$. Even so, the visibility does not match the expected $50\%$ value as can be observed from Fig. \ref{fig:EffectivenessRegion}, an effect we attribute to experimental misalignment up to the detector's sensitivity and polarization fluctuations inside the fiber link which, once again, are in accordance with the results of Fig. \ref{fig:LaserCharacterizationSetup}-b. We experimentally extended the average number of photons per time interval to the limit of the SPAD ($\mu \approx 10$ for $w_g=4$ ns, the detector's gate width) and observed that, even though the visibility starts to decay abruptly, the interference pattern is still reproduced in the interferometer. The results of Fig. \ref{fig:EffectivenessRegion} therefore show that the upper limitation of the Few-Photon FTS can be extended beyond the model of \cite{ThiagoJOSAB2015} at the cost of loss in visibility.

At the same time, a high power local oscillator allows the classical technique to read optical signals down to tens of photons per nanosecond on the nano-watt level. The heterodyne beat between optical signals results in an oscillatory term with electrical power directly proportional to the product of the optical signals, $P_{beat}^{elec} \propto P_{ref}^{opt}P_{test}^{opt}$. This means that low-power test signals can, in principle, be read with a sufficiently high-power reference laser. The power range of the classical technique is therefore upper- and down-limited by the saturation of the detector and the noise floor of the detection system, respectively. The Relative Intensity Noise is also a concern and can limit the SNR of the classical technique \cite{AgrawalBOOK2002}.

The Few-Photon FTS technique allows for the characterization of states with average number of photons per time interval down to the limit imposed by the dark count probability of the SPADs. The average number of photons per time interval in this case corresponds to $-118$ dBm in the Telecom C-band -- computed as $P = \mu h c /\left(\lambda w_g\right)$, where $P$ is the optical power, $h$ is Planck's constant, $c$ is the speed of light in vacuum, and $\lambda$ is the wavelength in vacuum. The Few-Photon FTS is therefore capable of stretching the achievable region of spectral characterization of optical sources, from the nano-watt level of the classical technique, down to the femto-watt level.

In conclusion, we have derived the mathematical model which permits one to associate the Fourier Transform of the interferogram of the two-photon interference in a Hong-Ou-Mandel interferometer to the spectral shape of the interfering wave-packets. We presented the characterization of weak coherent states from a faint laser source using the Few-Photon Fourier Transform Spectroscopy method and showed that the experimental results agree with the model so frequency-displaced and broadened-linewidth laser sources in the weak regime can be characterized with the proposed technique. The heterodyne beat of bright versions of the laser sources is compared in each case and the results show that the techniques are equivalent. Even though the compromise between the technique's resolution and range and its time of measurement limits the contribution from the whole spectrum of the sources, the Few-Photon FTS is shown to be valid for a range of optical intensity not covered by the classical spectroscopy method.

\begin{flushleft}
\textbf{Funding.} Work partially supported by CNPq and FAPERJ.
\end{flushleft}

\bibliographystyle{IEEEtran}
\bibliography{References}
\vfill

\end{document}